\documentclass[aps,prl,showpacs,twocolumn,amsmath,amssymb,
superscriptaddress]{revtex4}
\usepackage{graphicx}
\usepackage{hyperref}
\begin{document}
\title{Model of the Influence of an External Magnetic Field on the Gain of
Terahertz Radiation from Semiconductor Superlattices}
\author{Timo Hyart}
\affiliation{Department of Physics, P. O. Box 3000, FI-90014
University of Oulu, Oulu, Finland}

\author{Jussi Mattas}
\affiliation{Department of Physics, P. O. Box 3000, FI-90014
University of Oulu, Oulu, Finland}

\author{Kirill N. Alekseev}
\affiliation{Department of
Physics, P. O. Box 3000, FI-90014 University of Oulu, Oulu, Finland}
\affiliation{Department of Physics, Loughborough University,
Loughborough, LE11 3TU, United Kingdom}


\begin{abstract}
We theoretically analyze the influence of magnetic field on
small-signal absorption and gain in a superlattice. We predict a
very large and tunable THz gain due to nonlinear cyclotron
oscillations in crossed electric and magnetic fields. In contrast to
Bloch gain, here the superlattice is in an electrically stable
state. We also find that THz Bloch gain can be significantly
enhanced with a perpendicular magnetic field. If the magnetic field
is tilted with respect to the superlattice axis, the usually
unstable Bloch gain profile becomes stable in the vicinity of
Stark-cyclotron resonances.
\end{abstract}

\maketitle

Semiconductor superlattice (SL) is a model system for a wealth of
fundamental phenomena resulting from the wave-nature of charge
carriers \cite{Esaki, Bassrev, wackerrew}. Interesting examples of
such phenomena are Bloch oscillations in dc biased SLs
\cite{BOsSLs}, Shapiro-like steps \cite{unterrainer} and parametric
resonance \cite{hyart-param} in time-dependent electric fields,
coherent Hall effect in crossed electric and magnetic fields
\cite{crossedfield-tran-exp} and Stark-cyclotron resonances in
tilted magnetic field \cite{tilted-theory-exp}. The semiclassical
theory predicts that electrons performing Bloch oscillations in the
presence of weak dissipation can potentially provide THz Bloch gain
\cite{KSS}. The Bloch gain profile, which is shaped as a familiar
dispersion curve, is not limited to SLs. It was recently predicted
\cite{Willenberg} and observed \cite{Faist} in intersubband
transitions of quantum cascade lasers, but dispersive gain profiles
have been also found in the microwave responses of Josephson
junctions \cite{JJs} and Thim amplifiers \cite{Thim}. Here we
consider the Bloch gain in its traditional meaning as
an effect occurring due to Bloch oscillations in a single energy
band. In SLs, the realization of Bloch oscillator is a long-standing
problem due to the instability of a homogeneous electric field in
conditions of negative differential conductivity (NDC)
\cite{Ridley63-ignatov87}. This electric instability results in a
formation of electric domains in long SLs, which are destructive for
the THz gain.

A considerable amount of theoretical and experimental activities
have been devoted to the investigations of transient oscillations
and voltage-current (VI) characteristics in the presence of magnetic
field \cite{Bass81, Bassrev, Feil06, Polyanovskii80--theory,
crossed-fields-experiments, crossedfield-tran-exp,
tilted-theory-exp}, but the response to a time-dependent field is
mostly unexplored. In this letter, we focus on the influence of the
magnetic field on the small-signal absorption and gain. We find that
absorption and gain profiles in crossed electric and magnetic fields
have different characteristic shapes in the Bloch-like and
cyclotron-like regimes of ballistic motion. We predict a very large
and tunable THz gain due to nonlinear cyclotron oscillations in a
single energy band. The cyclotron gain takes place in conditions of
positive differential conductivity (PDC) and therefore it is stable
against space-charge fluctuations. We also demonstrate that Bloch
gain can be significantly enhanced by the perpendicular magnetic
field. Moreover, the usually unstable Bloch gain profile becomes
stable in the vicinity of Stark-cyclotron resonances if the magnetic
field is tilted with respect to the SL axis.

We consider a SL under an action of a static magnetic field
$\mathbf{B}$ in arbitrary direction and an electric field
$E(t)=E_{dc}+E_\omega \cos\omega t$ in a SL direction, which is
chosen to be the $x$-direction. Here $E_\omega \cos\omega t$ is a
weak probe field with a frequency $\omega$ fixed by an external
circuit (external cavity). We consider the electron transport in a
single miniband with the standard tight-binding dispersion relation
\begin{equation}
\varepsilon({\bf k})=-(\Delta/2) \cos k_x
 d+\hbar^2(k_y^2+k_z^2)/2m,
\end{equation}
where $\varepsilon(\mathbf{k})$ is the electron energy, $\mathbf{k}$
is its quasimomentum, $\Delta$ is the miniband width, $d$ is the SL
period and $m$ is the effective mass in $y$- and $z$-directions. We
use the semiclassical approach based on the Boltzmann equation in
the relaxation time $\tau$ approximation
\begin{equation}
\frac{\partial f}{\partial t}+\frac{1}{\hbar} \bigg[e \mathbf{E}
+e\mathbf{v}\times\mathbf{B}\bigg] \cdot \frac{\partial f}{\partial
\mathbf{k}}=-\frac{f-f_{eq}}{\tau}, \label{Boltz}
\end{equation}
where $v_i(k_i)=\hbar^{-1}\partial \varepsilon(\mathbf{k})/\partial
k_i$ ($i=x,y,z$) are the velocity components and
$f_{eq}(\mathbf{k})$ is the Fermi distribution \cite{Bassrev,
wackerrew}. By solving Eq.~(\ref{Boltz}), we find the stationary
time-dependent current after transient
\begin{equation}
\mathbf{j}(t)=\frac{2e}{(2\pi)^3 \tau} \int d^3k \
f_{eq}(\mathbf{k})
 \int_{-\infty}^t ds \ e^{-(t-s)/\tau}
 \mathbf{v}(\mathbf{k}_s^t), \label{curdens}
\end{equation}
where the prefactor takes into account the density of states
\cite{wackerrew}, $k_x$ is integrated over the Brillouin zone and
the integration limits for $k_y$ and $k_z$ are $\pm\infty$. Here
$\mathbf{k}_s^t$ is a ballistic trajectory governed by the equations
\begin{equation}
\frac{d \mathbf{k}_s^t}{dt}=\frac{1}{\hbar} \bigg[e \mathbf{E} (t)
+e\mathbf{v}(\mathbf{k}_s^t)\times\mathbf{B}\bigg], \hspace{0.5
cm} \mathbf{k}_s^s=\mathbf{k}. \label{ballistic}
\end{equation}
The second equation in (\ref{ballistic}) means that the
quasimomentum at $t=s$ is $\mathbf{k}$. We derived the solution of
Boltzmann equation (\ref{curdens}) by using both a generalization of
the technique based on time-evolution operator \cite{Bass81} and the
method of characteristics \cite{MacCallum}. The approach based on
this solution is most powerful in the limits of low carrier density
$N$ and low temperature. In this case
\begin{equation}
f_{eq} \approx 4 N \pi^3 \delta(\mathbf{k}) \label{lowT}
\end{equation}
and the current (\ref{curdens}) is determined by the ballistic
trajectories starting at $\mathbf{k}=\mathbf{0}$.

\begin{figure}
\includegraphics[width=0.49\columnwidth]{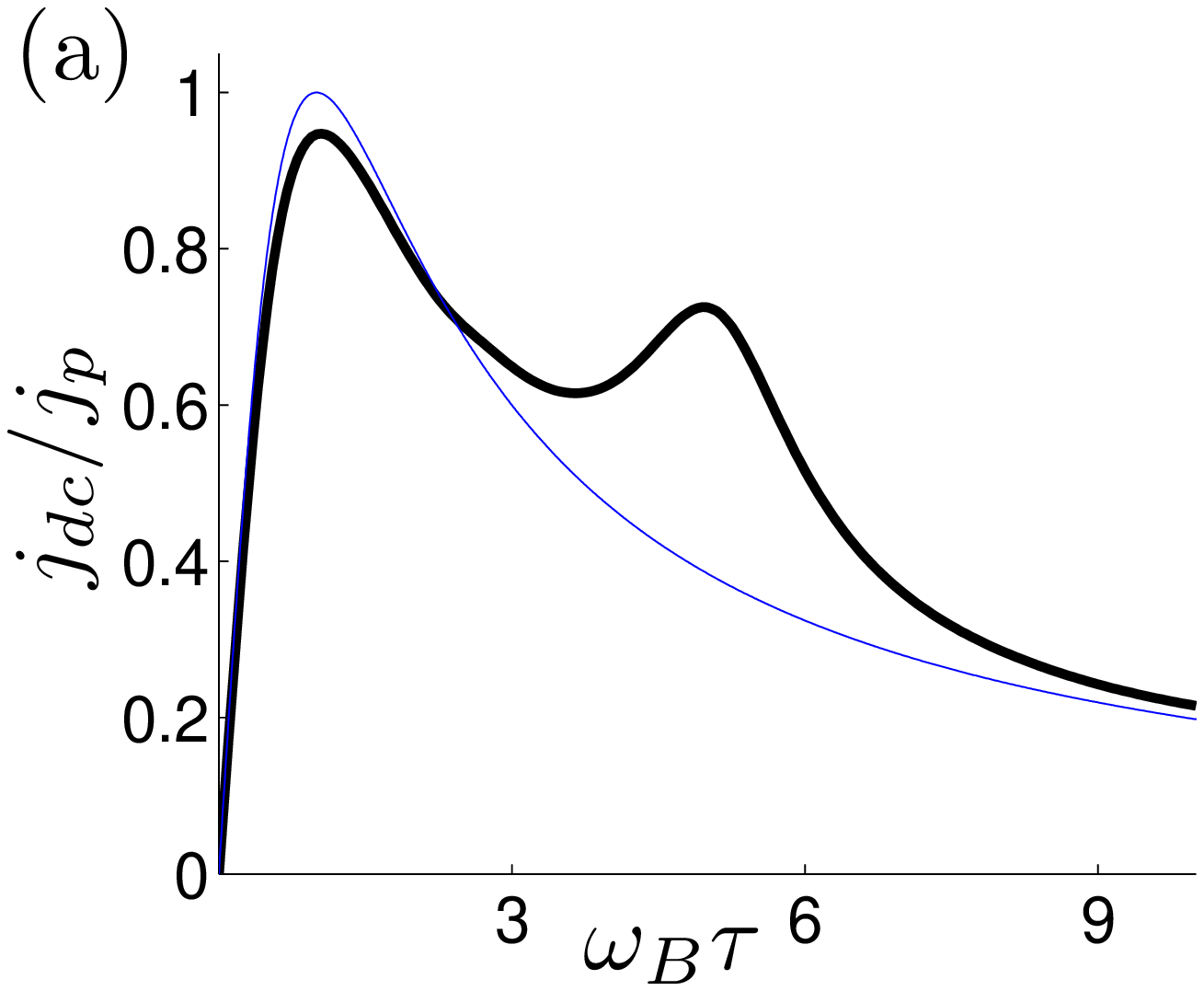}
\includegraphics[width=0.49\columnwidth]{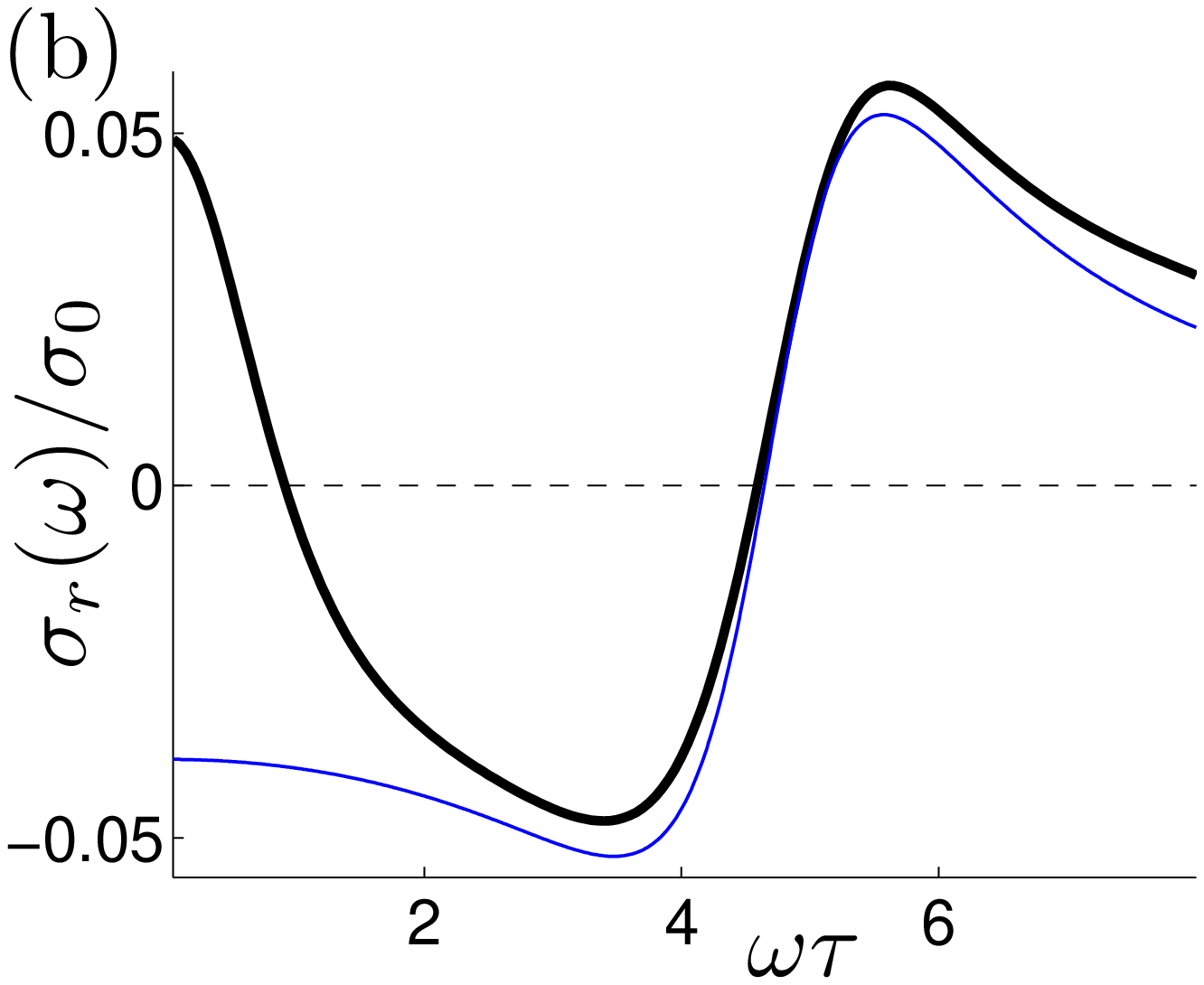}
\caption{(color online). (a) VI characteristic in tilted magnetic field, calculated using
Eqs.~(\ref{curdens})-(\ref{lowT}), for $\omega_c \tau=2$ and
$\omega_{c \bot} \tau=5$ (black). The thin blue line shows the
Esaki-Tsu VI characteristic [Eq.~(\ref{eq:ET})]. (b) Gain profile
for the same magnetic field and dc bias $\omega_B \tau=4.75$
(black).  The thin blue line indicates the usual Bloch gain at
$\mathbf{B}=\mathbf{0}$ [Eqs.~(\ref{Tucker-formula}) and
(\ref{eq:ET})]. } \label{tiltedgain}
\end{figure}

The time-dependent current (\ref{curdens}) can be used to calculate
the real part of the dynamical conductivity $\sigma_r(\omega) \equiv
{\rm Re}[\sigma(\omega)]$, which determines the gain ($\sigma_r <0$)
and absorption ($\sigma_r > 0$). In the absence of magnetic field
$\sigma_r$ is defined by the Tucker formula \cite{tucker79,
wackerrew}
\begin{equation}
\sigma_r(\omega)=\frac{j_{dc}(eE_{dc}d+\hbar
\omega)-j_{dc}(eE_{dc} d-\hbar \omega)}{2 \hbar \omega}ed,
\label{Tucker-formula}
\end{equation}
where $j_{dc}(e E_{dc} d)$ is the Esaki-Tsu characteristic
\cite{Esaki},
\begin{equation}
\label{eq:ET} j_{dc}(e E_{dc} d)=j_p \frac{2 e E_{dc} d
\tau/\hbar}{1+(e E_{dc} d \tau/\hbar)^2},
\end{equation}
and $j_p$ is the peak current corresponding to the critical field
$E_{cr}=\hbar/ed\tau$. The Drude conductivity of the SL is
$\sigma_0=2j_{p}/E_{cr}$. If $E_{dc} > E_{cr}$,
Eq.~(\ref{Tucker-formula}) describes the dispersive Bloch gain
profile [blue line in Fig.~\ref{tiltedgain} (b)] with a crossover
from gain to loss at the resonance $\omega \approx \omega_B$, where
$\omega_B=eE_{dc}d/\hbar$ is the Bloch frequency. As directly
follows from Eqs.~(\ref{Tucker-formula}) and (\ref{eq:ET}), the
upper limit for this Bloch gain is $\min [\sigma_r(\omega)]
=-\sigma_0/8$.

\begin{figure}
\includegraphics[width=0.49\columnwidth]{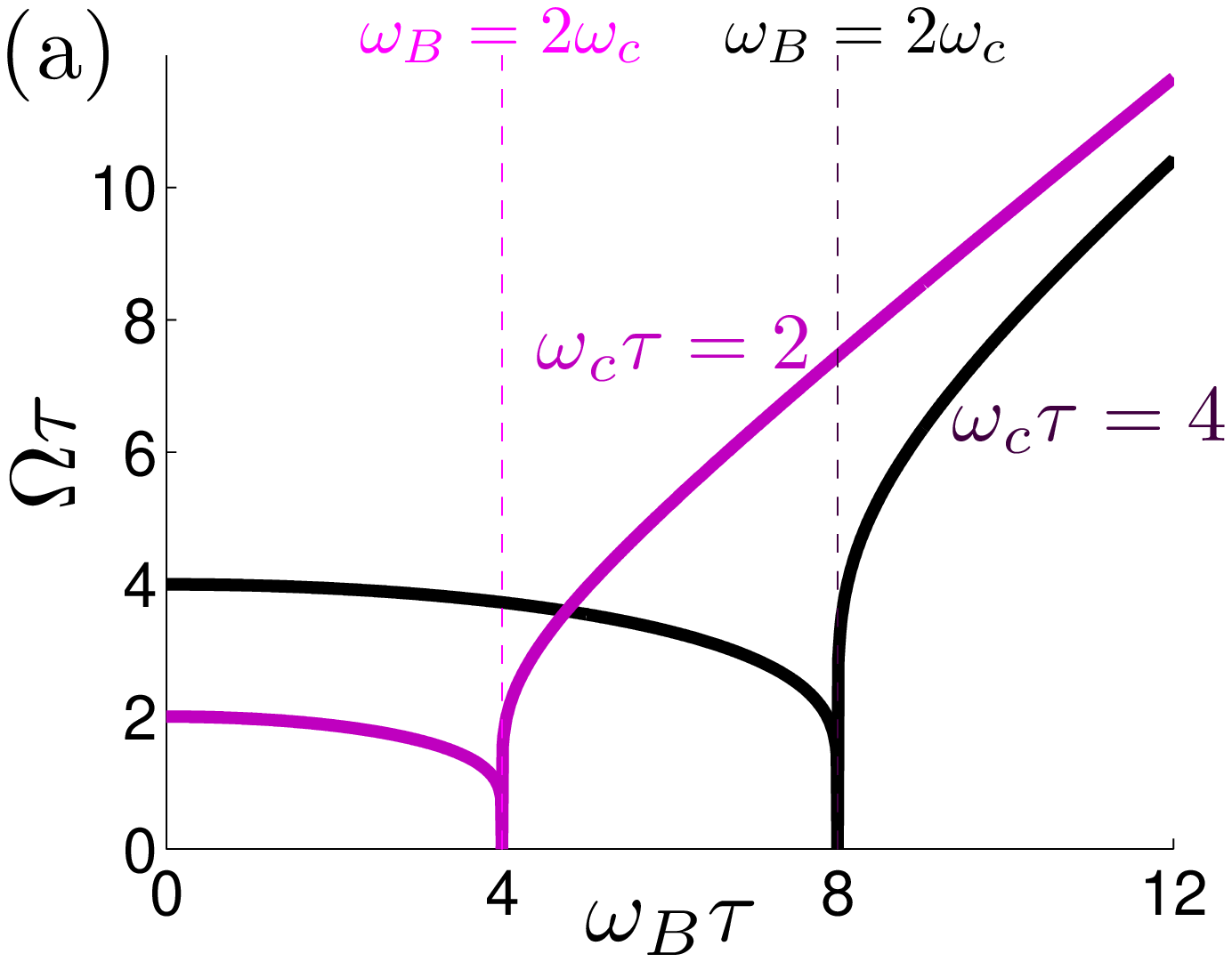}
\includegraphics[width=0.49\columnwidth]{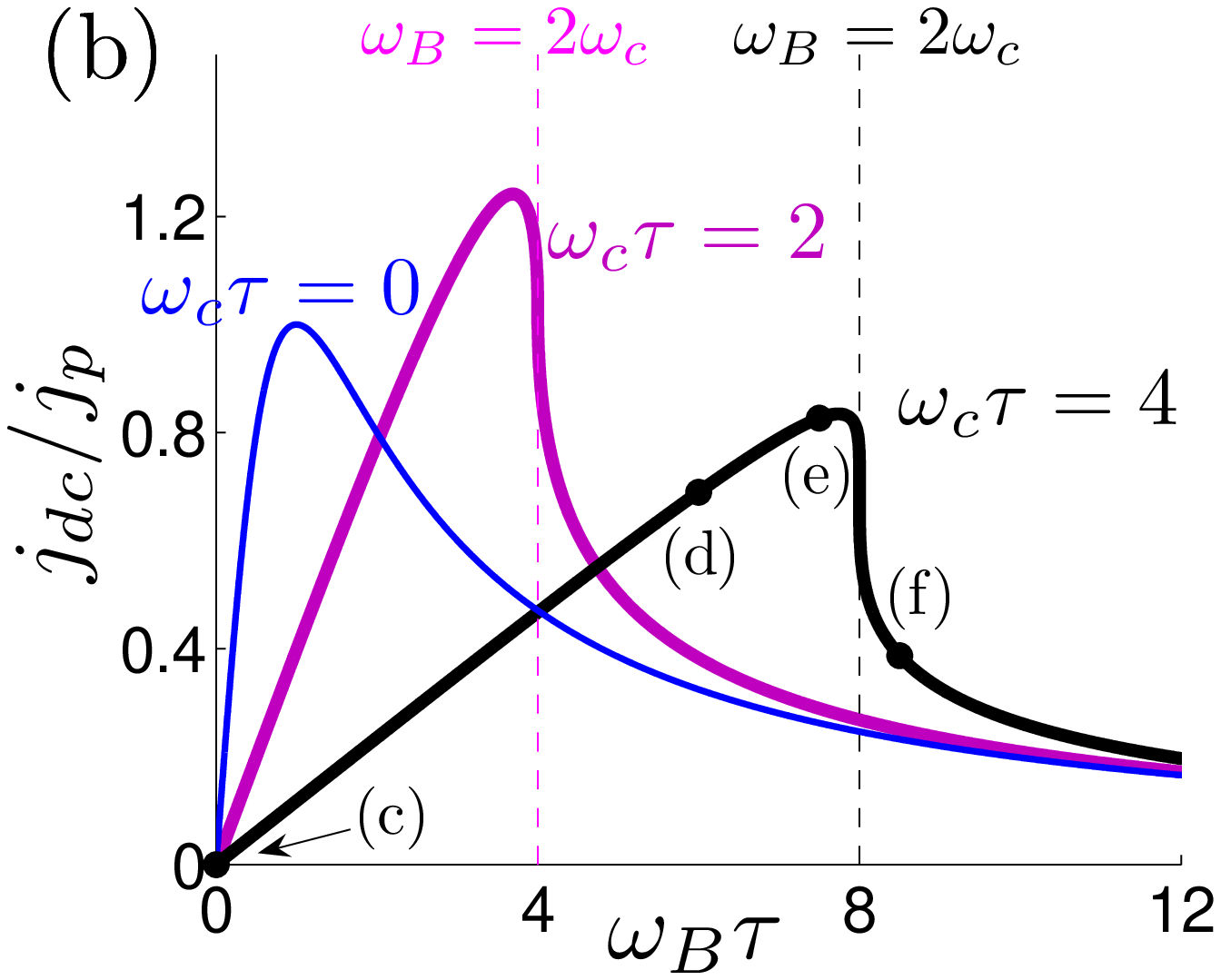}
\includegraphics[width=0.49\columnwidth]{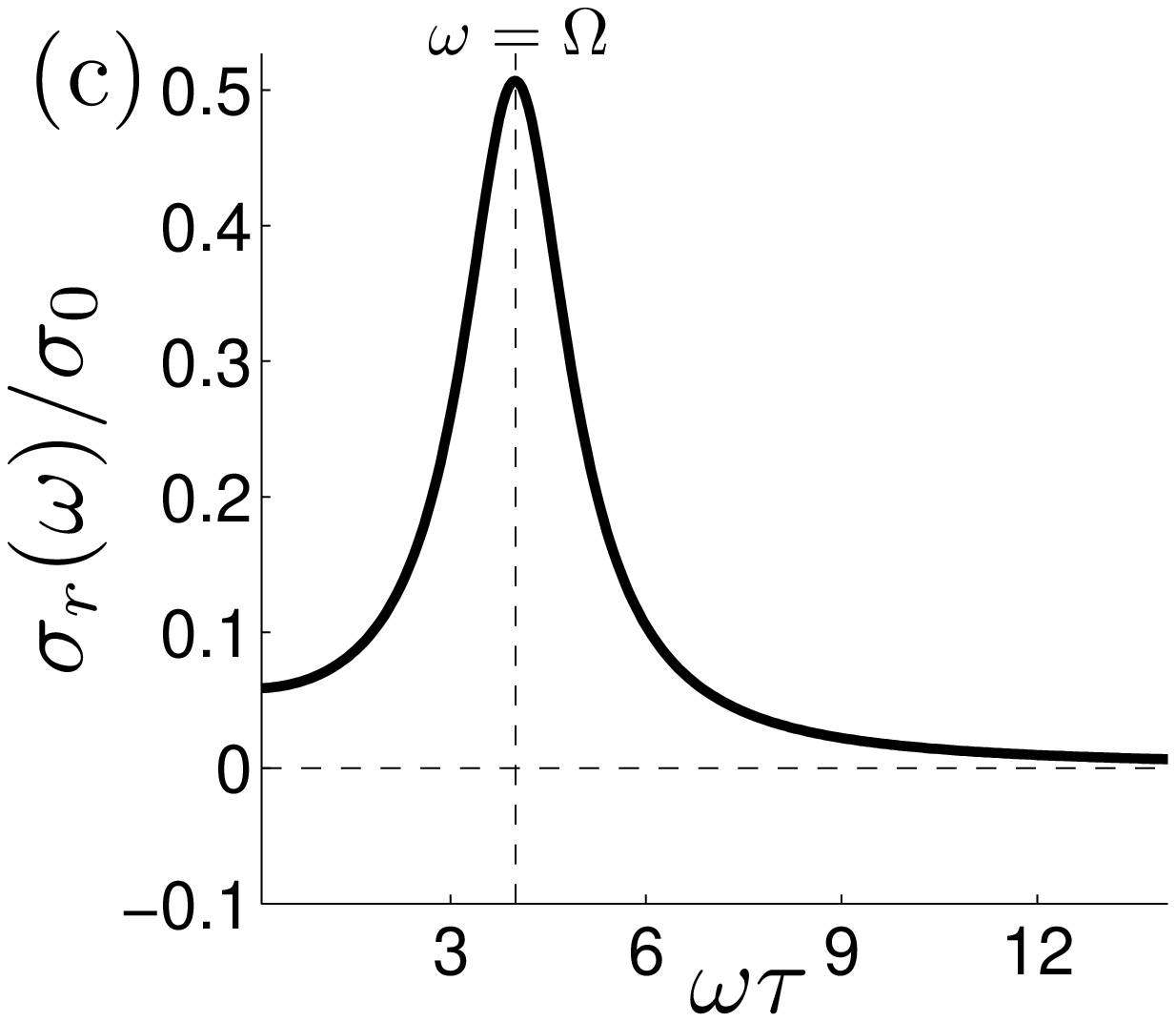}
\includegraphics[width=0.49\columnwidth]{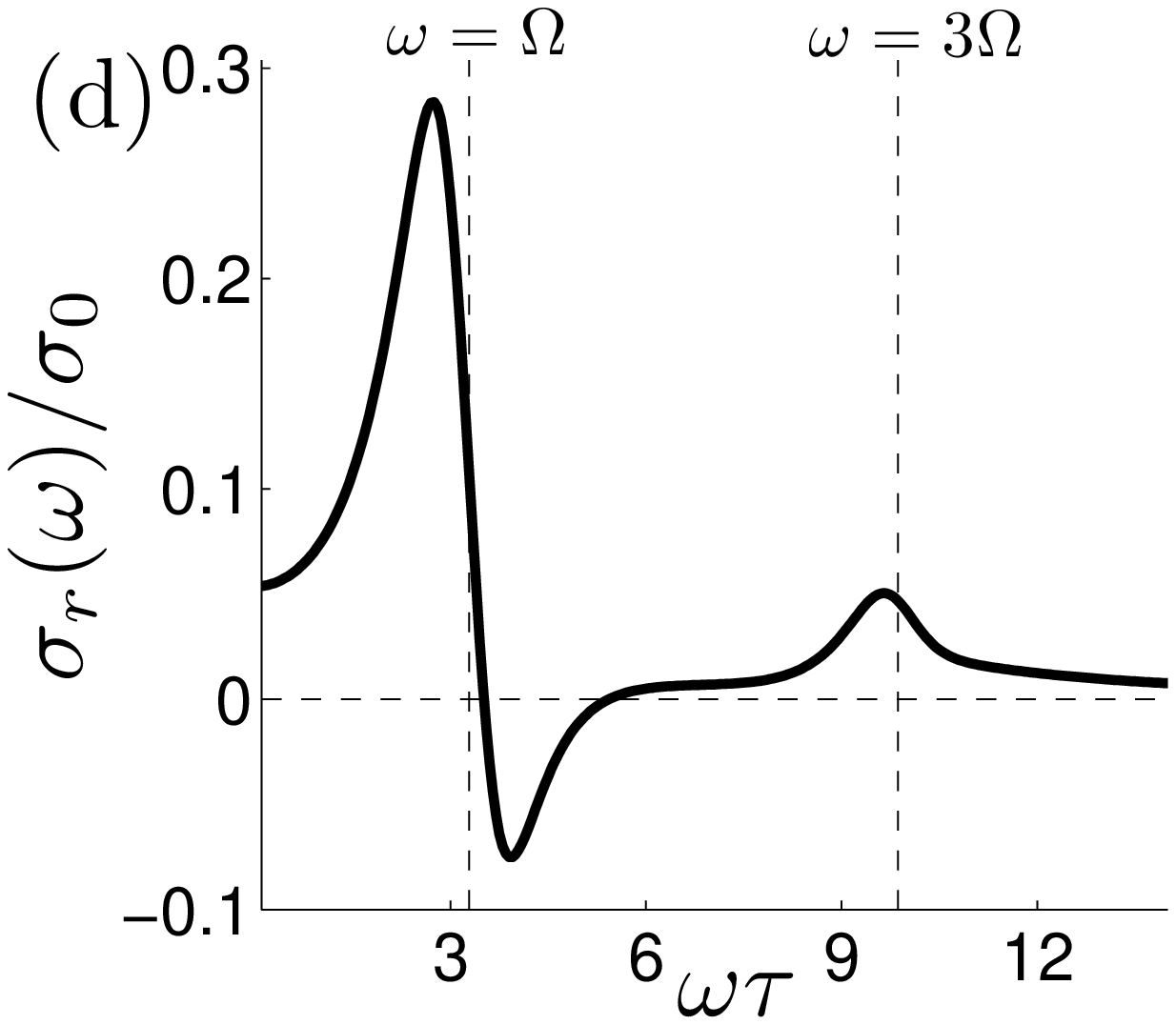}
\includegraphics[width=0.49\columnwidth]{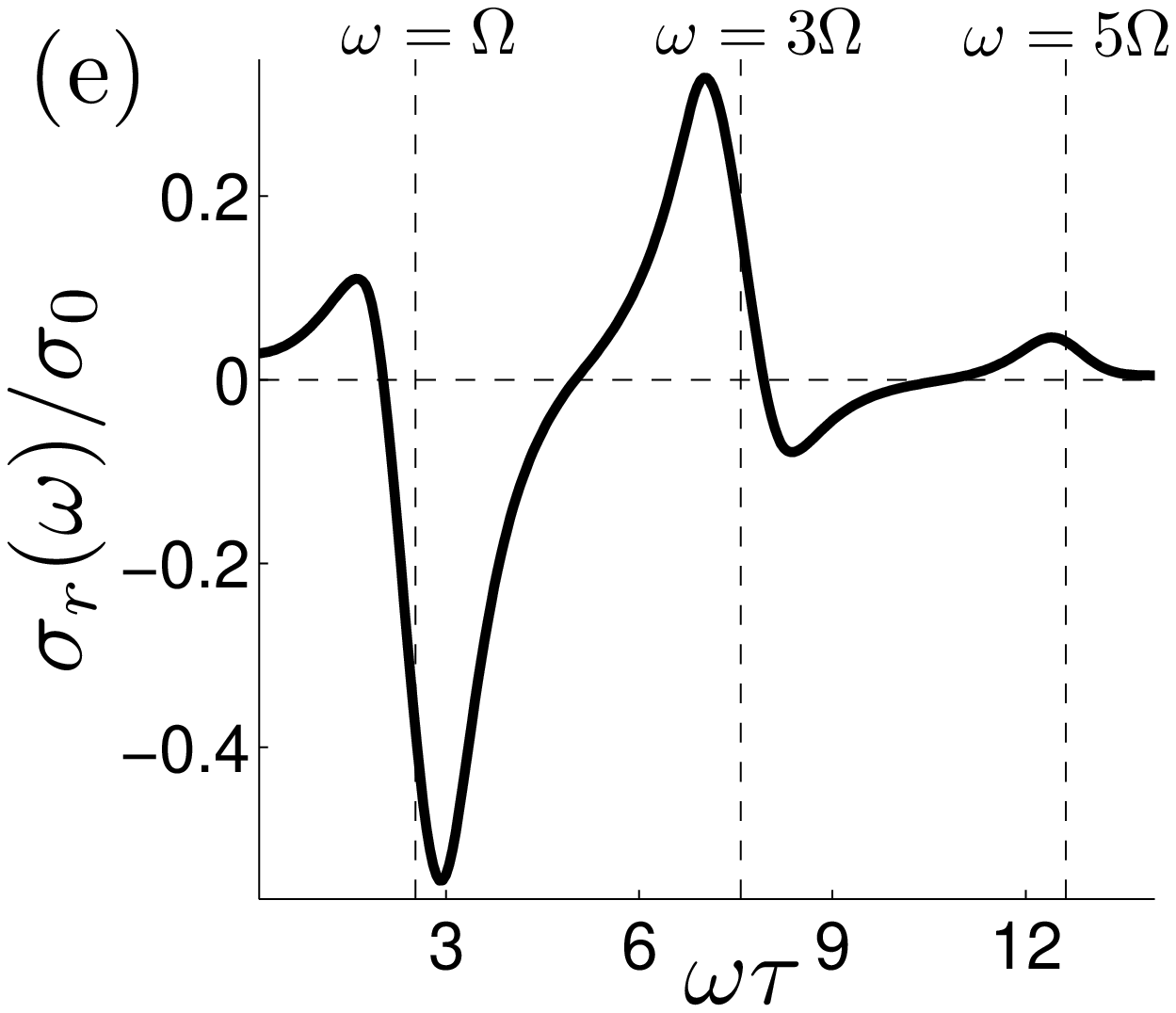}
\includegraphics[width=0.49\columnwidth]{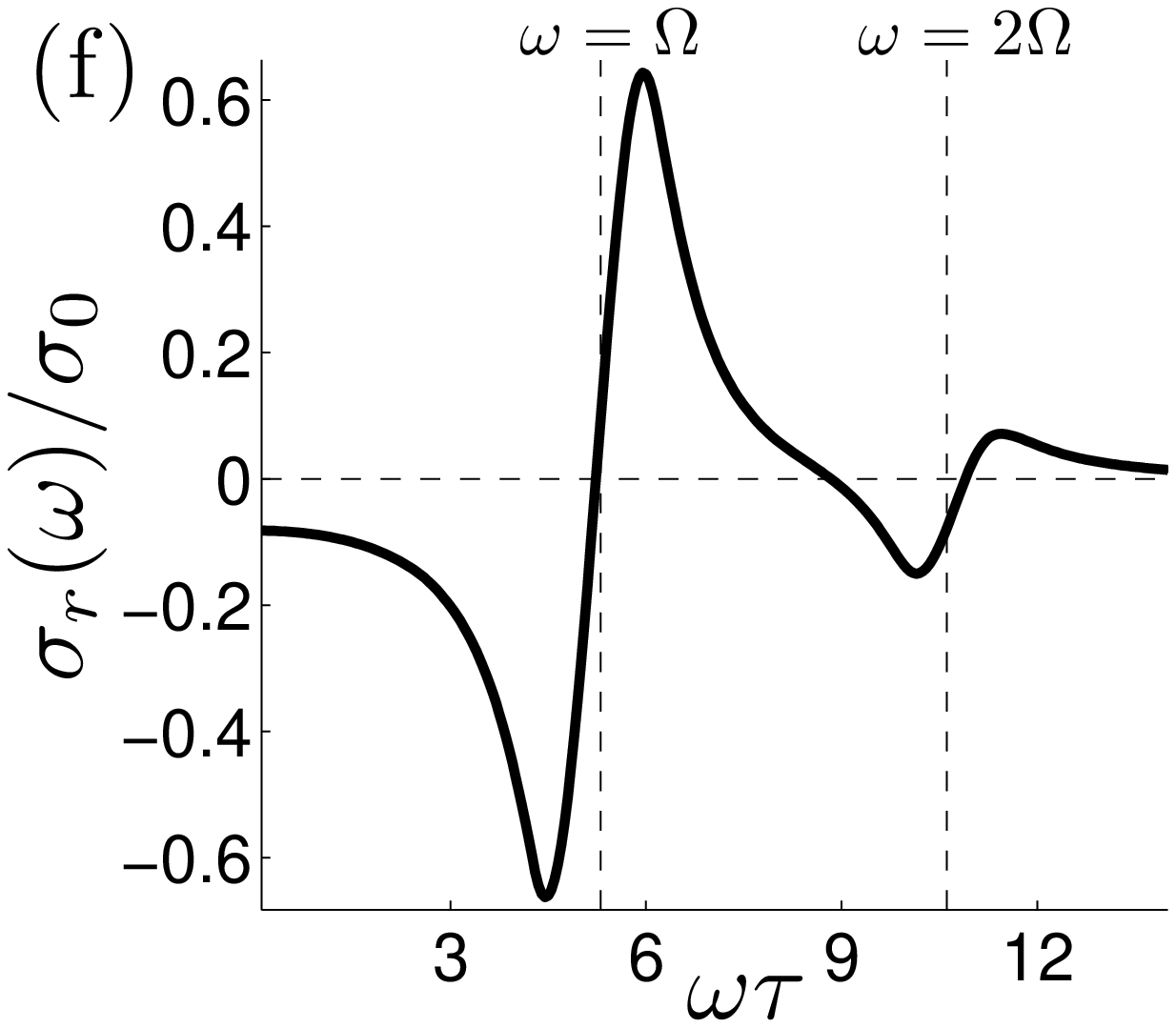}
\caption{(color online). (a) Frequency of the nonlinear oscillations
$\Omega$ [Eq.~(\ref{resfreq})] as a function of $\omega_B$ for
$\omega_c \tau=:2,4$. The dashed lines separate the cyclotron- and
Bloch-like regimes of motion. (b) VI characteristics for $\omega_c
\tau=:0, 2,4$. (c)-(f) Absorption and gain profiles for $\omega_c
\tau=4$ and different values of electric field $\omega_B \tau =:$
$0$ (c), $6$ (d), $7.5$ (e) and $8.5$ (f).} \label{gain-vernissage}
\end{figure}

We now turn to the consideration of electron dynamics in crossed
electric and magnetic fields $\mathbf{B}=(0,0,B_z)$. The ballistic
electron dynamics under static fields is determined by the pendulum
equation, which follows from Eq.~(\ref{ballistic}) \cite{Bassrev,
Polyanovskii80--theory}. If the electrons start at the bottom of the
miniband, the frequency of their nonlinear oscillations is
\begin{eqnarray}
\Omega&=&\left\{
\begin{array}{rlc}
 \pi\omega_c/2K(\omega_B/2\omega_c),  &  \omega_B < 2\omega_c \\
 \pi\omega_B/2 K(2 \omega_c/\omega_B),  & \omega_B > 2\omega_c,
\end{array}
\right. \label{resfreq}
\end{eqnarray}
where $\omega_c=eB_z/\sqrt{m_x m}$ and $m_x=2\hbar^2/\Delta d^2$ are
the cyclotron frequency and the effective mass at the bottom of the
miniband, respectively. Here $K(k)$ is the complete elliptic
integral of the first kind as a function of elliptic modulus. There
exist two distinct regimes of motion: cyclotron-like oscillations
for large magnetic field $\omega_B < 2\omega_c$ (oscillations of
pendulum) and Bloch-like oscillations for dominating electric field
$\omega_B > 2\omega_c$ (rotations of pendulum). As can be seen from
Fig.~\ref{gain-vernissage} (a), the frequency $\Omega$ is tunable by
variation of the magnetic and electric fields. If $\omega_B \gg
\omega_c$, we get from Eq.~(\ref{resfreq}) that $\Omega=\omega_B$.
In the opposite limit $\omega_B \ll \omega_c$ we have
$\Omega=\omega_c$. Close to the separatrix $\omega_B=2\omega_c$
oscillations are strongly nonlinear and anharmonic. All harmonics of
$\Omega$ are present in the Bloch-like regime whereas only odd
harmonics exist in the cyclotron-like regime. Such kind of transient
oscillations have been directly observed in the experiments
\cite{crossedfield-tran-exp}.

\begin{figure}
\includegraphics[width=0.49\columnwidth]{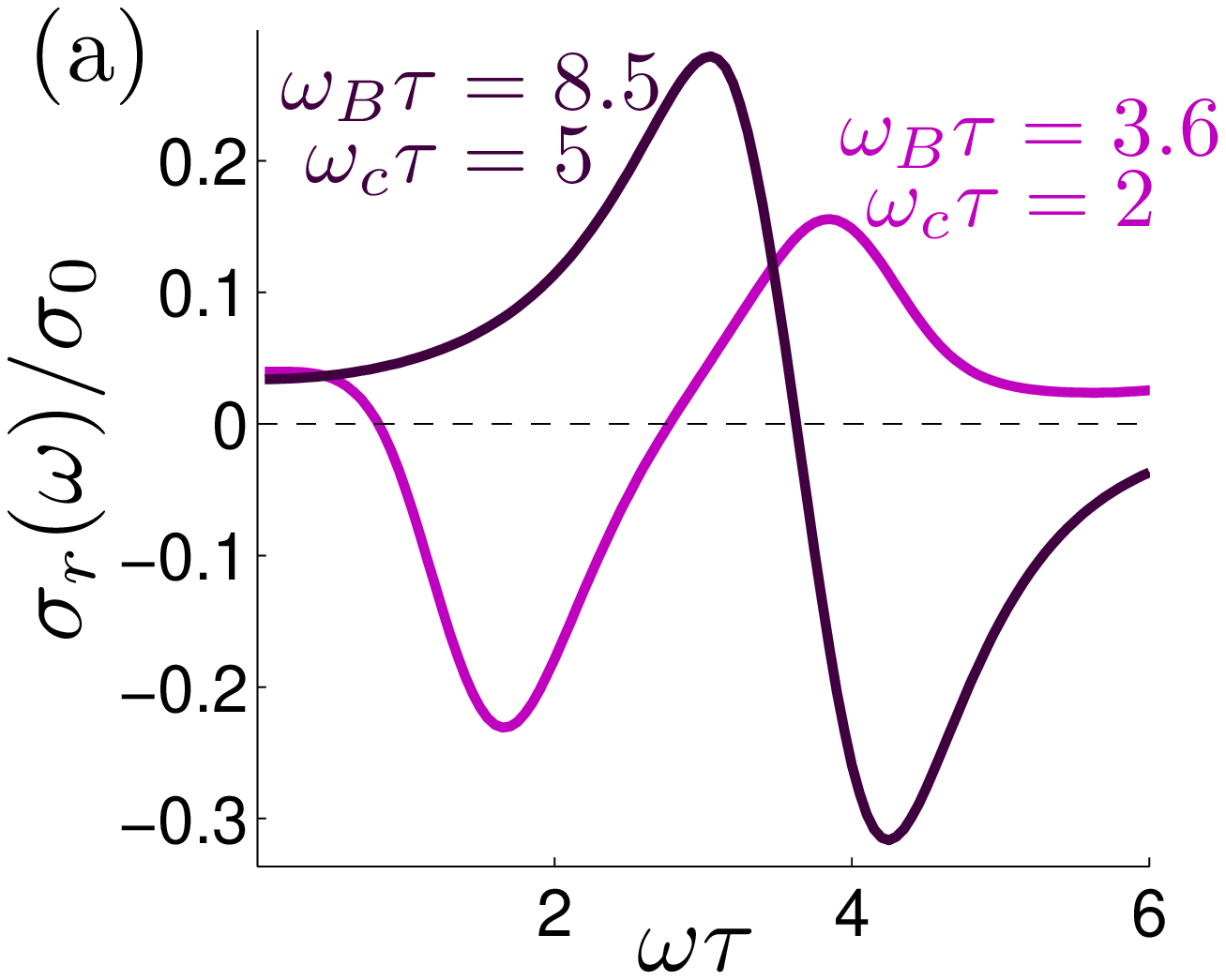}
\includegraphics[width=0.49\columnwidth]{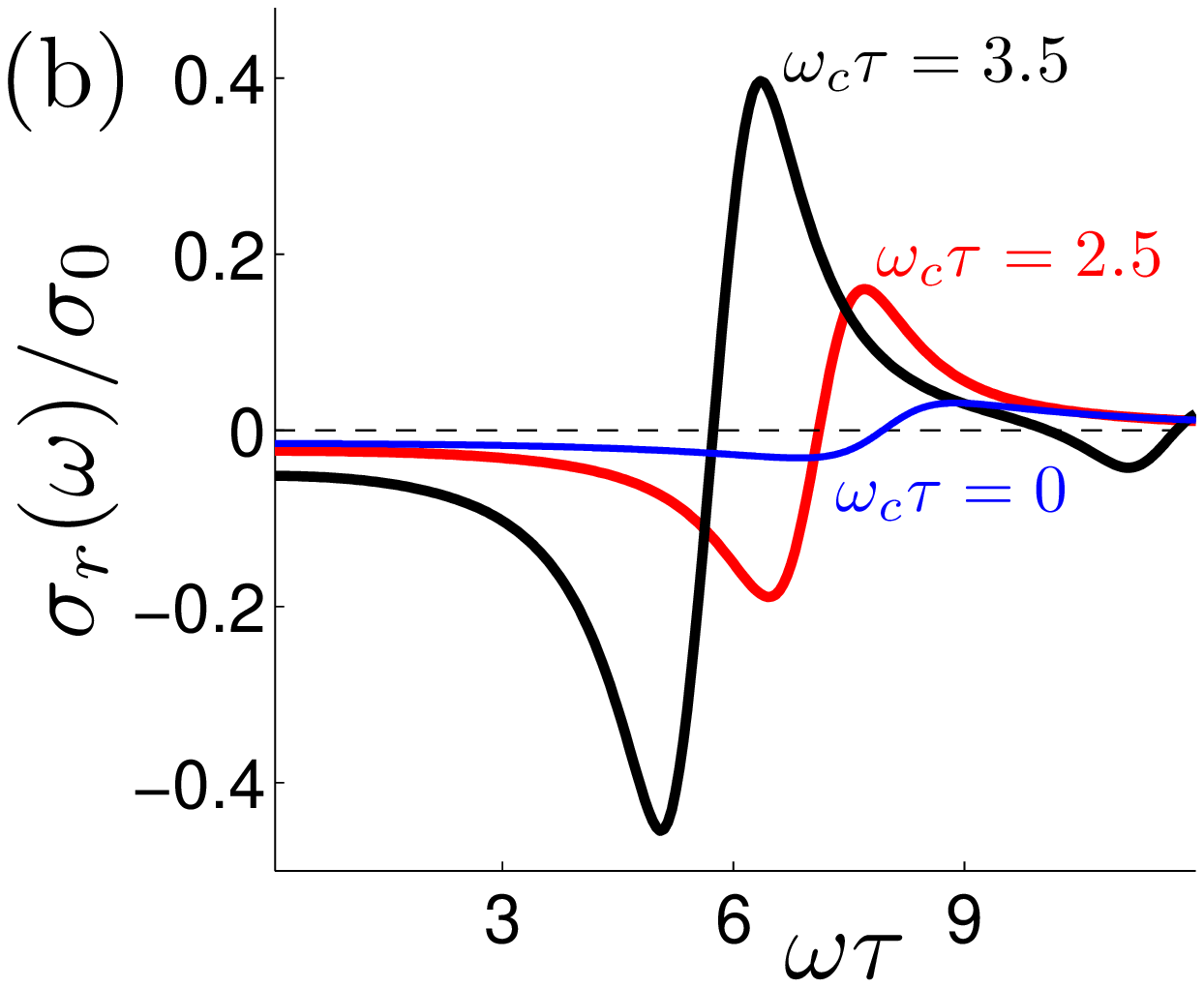}
\caption{(color online). (a) Cyclotron gain profiles for $\omega_B
\tau=3.6$, $\omega_c \tau=2$ and $\omega_B \tau=8.5$, $\omega_c
\tau=5$. (b) Bloch gain profiles for $\omega_B \tau=8$ and different
values of the magnetic field $\omega_c \tau =:0, 2.5, 3.5$. The gain
profiles were calculated using Eqs.~(\ref{curdens})-(\ref{lowT}).}
\label{enhBG-cyclogain}
\end{figure}

In the presence of scattering these oscillations determine the dc
current density at low temperatures, as can be directly seen from
Eqs.~(\ref{curdens})-(\ref{lowT}) \cite{Polyanovskii80--theory}. The
resulting VI characteristics for different values of the magnetic
field are shown in Fig.~\ref{gain-vernissage} (b). The transition
from oscillatory to rotational motion at the separatrix
$\omega_B=2\omega_c$ manifests itself as a very abrupt and strong
suppression of the dc current \cite{Polyanovskii80--theory,
crossed-fields-experiments, Feil06}. In the cyclotron-like regime,
there are no Bragg reflections and therefore it corresponds the PDC
part of the VI characteristic. On the other hand, the Bragg
reflections in the Bloch-like regime result in NDC at the operation
point. We find that the response of the miniband electrons to a weak
ac field also shows clear signatures of the different types of
ballistic motion. Figs.~\ref{gain-vernissage} (c)-(f) show the
absorption and gain profiles, calculated using
Eqs.~(\ref{curdens})-(\ref{lowT}), for different values of
$\omega_c$ and $\omega_B$. The field strengths are chosen in such a
way that the electrons perform several cycles of oscillations
between the scattering events $\Omega \tau > 1$. We see that the
gain and absorption profiles in the different regimes of
oscillations have their own characteristic shapes. If $\omega_c \gg
\omega_B$, electrons are restricted to the bottom of the miniband,
and we obtain a familiar Lorentzian absorption profile of a harmonic
oscillator, which is centered at $\omega \approx \Omega$
[Fig.~\ref{gain-vernissage} (c)]. In this case $\sigma_r$ is always
positive and an amplification of probe field is not possible. By
increasing the electric field $E_{dc}$ we obtain a completely
different situation, where high-frequency gain can appear due to
nonlinear cyclotron oscillations [Fig.~\ref{gain-vernissage} (d),
(e)]. The characteristic shape of the gain profile turns out to be
an inverse of the usual dispersive Bloch gain profile [\textit{cf.}
Fig.~\ref{gain-vernissage} (d) and Fig.~\ref{enhBG-cyclogain} (b)].
In this gain profile the frequency of the nonlinear cyclotron
oscillations $\Omega$ determines the position of a resonant
crossover from loss to gain. When we are approaching the separatrix
$\omega_B=2\omega_c$, we see that replicas of this inverted
dispersive gain profile appear at odd harmonics of $\Omega$
corresponding to the anharmonicity of the ballistic oscillations
[Fig.~\ref{gain-vernissage} (e)]. With a further increase of
$E_{dc}$ we arrive to Bloch-like regime $\omega_B > 2\omega_c$. Here
we always have a dispersive Bloch gain profile with a crossover
frequency $\Omega$ [Fig.~\ref{gain-vernissage} (f)].
Replicas of this gain profile can now appear at all harmonics of $\Omega$.

Our results show that both the cyclotron gain and the Bloch-like
gain have such attractive properties as tunability and extremely
large magnitudes of the gain, which can be utilized in an operation
of THz oscillators and amplifiers.  The gain profiles in both
regimes of motion are tunable by changing magnetic and electric
fields. In the Bloch-like regime, the gain profile can be controlled
more easily with a variation of the electric field, because $\Omega$
depends strongly on $\omega_B$ [see Fig.~\ref{gain-vernissage} (a)].
In this regime $\Omega$ is sensitive to the magnetic field only near
the separatrix.  On the other hand, since the cyclotron gain exists
only near the separatrix, it is tunable by simultaneous variation of
the electric and magnetic fields as shown in
Fig.~\ref{enhBG-cyclogain} (a). We see that large cyclotron gain can
be obtained at frequencies $\omega \tau \geq 1$ if the cyclotron
frequency is somewhat larger than the scattering rate $\omega_c \tau
\geq 2$ [Fig.~\ref{enhBG-cyclogain} (a)]. In typical SLs $\tau
\approx 200$ fs so that $\omega \tau=1$ corresponds to the frequency
$\omega/2\pi=0.8$ THz. Therefore the cyclotron gain, similarly as
the Bloch gain, is ideal for amplifiers and oscillators operating at
THz frequencies.

We turn to the important issues of the magnitudes and the origin of
the high-frequency gain. Fig.~\ref{enhBG-cyclogain} (b) shows the
Bloch gain profiles for fixed $\omega_B$ and different values of
$\omega_c$. We see that by increasing the magnetic field the gain
increases roughly by an order of magnitude in comparison with the
usual Bloch gain at $\mathbf{B}=\mathbf{0}$. Similarly, the THz
cyclotron gain near the separatrix [Fig.~\ref{gain-vernissage} (e)]
is also an order of magnitude larger than the usual Bloch gain. For
typical SL parameters $d=6$ nm, $\Delta=60$ meV and $m=0.067 m_e$,
we obtain that $\omega_c \tau=1$ and $\omega_B \tau=1$ correspond to
magnetic field $B=2$ T and electric field $E_{dc} = 5.5$ kV/cm,
respectively. Thus, already rather weak electric and magnetic fields
can provide strong THz gain [Fig.~\ref{enhBG-cyclogain} (a)]. The
magnitude of the gain $\alpha$ in units $\rm{cm}^{-1}$ is related to
the dynamical conductivity as $\alpha=\alpha_0 (\sigma_r/\sigma_0)$,
where at low temperatures $\alpha_0=Ne^2 \tau/c \sqrt{\epsilon}
\epsilon_0 m_x$. For moderate doping $N= 10^{16}$ cm$^{-3}$ and
relative permittivity $\varepsilon=13$ (GaAs), we obtain $\alpha_0
\approx 833.5 \ \textrm{cm}^{-1}$. Using this value of $\alpha_0$,
it is easy to estimate all magnitudes of the gain in
Figs.~\ref{gain-vernissage} and \ref{enhBG-cyclogain}. In the
vicinity of the separatrix THz gain can have unprecedented values
$\alpha \approx 500 \ \rm{cm}^{-1}$.

Although the response to the ac field results from quite complicated
nonlinear electron dynamics, we can identify an important role of
the separatrix in the origin of this large THz gain. In the
Bloch-like regime electrons periodically visit the upper part of the
miniband and undergo Bragg reflections at the boundary of the
Brillouin zone. In contrast there are no Bragg reflections in the
cyclotron-like regime. However, a strong enough electric field
$E_{dc}$ periodically brings the carriers to the upper part of the
miniband where their effective masses are negative \cite{Esaki,KSS}.
We attribute the cyclotron gain to these nonclassical cyclotron
oscillations in presence of weak dissipation (scattering). As the
electrons performing Bloch- and cyclotron-like oscillations are
spending more time in the upper part of the miniband the gain
profiles become more pronounced.  In the vicinity of the separatrix,
the electron trajectories stick a long time near the Brillouin zone
boundary (the hyperbolic point of the pendulum) resulting in a
significant enhancement of the THz gain [Figs.~\ref{gain-vernissage}
(e), \ref{enhBG-cyclogain} (b)].

Since in the cyclotron-like regime the SL operates in conditions of
PDC, it is inherently stable with respect to the space-charge
fluctuations. In the gain profiles this can be seen as positive
dynamical conductivity at low frequencies
[Fig.~\ref{gain-vernissage} (c)-(e), Fig.~\ref{enhBG-cyclogain}
(a)]. On the other hand, in the Bloch-like regime the gain always
extends to the low frequencies [Fig.~\ref{gain-vernissage} (f),
Fig.~\ref{enhBG-cyclogain} (b)] due to the operation in the NDC
state [Fig.~\ref{gain-vernissage} (b)]. This means that in the
Bloch-like regime the problem of electric instability exists
similarly as in the case of usual Bloch gain in SLs.
Several different approaches have been suggested in order to stabilize the Bloch gain,
including super-superlattice structures
\cite{savvidis}, quasistatic modulation of the bias \cite{Hyart09}
and 2D shunted surface SLs \cite{Feil06}. Here we demonstrate that
the electric instability can be circumvented by introducing an
additional magnetic field component in the SL direction, so that the
magnetic field becomes tilted with respect to the SL axis
$\mathbf{B}=(B_x,0,B_z)$. In this case, electrons perform cyclotron
oscillations in the plane perpendicular to the SL axis with the
frequency $\omega_{c\bot}=eB_x/m$. These in-plane cyclotron
oscillations are coupled to the Bloch oscillations via the
perpendicular magnetic field component $\omega_c$. If $\omega_B$ and
$\omega_{c\bot}$ are commensurate (Stark-cyclotron resonance), the
coupling results in delocalization of the electrons, which reveals
itself as additional resonant structures in the VI characteristics
\cite{tilted-theory-exp}. Fig.~\ref{tiltedgain} (a) demonstrates
such additional structure for a particular magnetic field. The
enhancement of the current at the Stark-cyclotron resonance
$\omega_B=\omega_{c \bot}$ resembles \cite{hyart08} the resonant
structures in VI characteristic induced by an auxiliary THz field
\cite{unterrainer}. By choosing the working point at the PDC part of
the peak, we find, using Eqs.~(\ref{curdens})-(\ref{lowT}), that the
gain profile close to the Bloch frequency $\omega \approx \omega_B$
has the usual shape of the dispersive Bloch gain, whereas the
dynamical conductivity at low frequencies is now positive
[Fig.~\ref{tiltedgain}(b)]. Interestingly, this stable gain profile
can be well approximated by a variant of the Tucker formula
(\ref{Tucker-formula}), in which $j_{dc}(eE_{dc}d)$ is now the
current density calculated in the presence of the magnetic field. We
should distinguish our results from the recent proposal of sub-THz
generation in SLs in tilted magnetic field \cite{Greenaway}. There
the moving charge domains are responsible for the current
oscillations, whereas in our case the THz gain occurs in the absence
of electric domains.

In summary, we found that the magnetic field significantly alters
the shapes of gain profiles and the magnitude of THz gain in SLs. We
described a novel type of large and tunable THz gain caused by
nonlinear cyclotron oscillations in the crossed electric and
magnetic fields. Since the operation point can be chosen at the PDC
part of the VI characteristic the old problem of space-charge
instability, which is typical for the Bloch gain in SL, does not
exist here. We also predicted an enhancement of the Bloch gain due
to nonlinear character of Bloch oscillations in the presence of a
perpendicular magnetic field. Finally, we demonstrated that in the
tilted magnetic field configuration the usual Bloch gain can be
realized in conditions of PDC.

We conclude with two remarks. First, by numerically solving the
Boltzmann equation (\ref{Boltz}), we observed that the cyclotron
gain decreases rapidly with increasing temperature.
It is typically smaller than the Bloch gain \cite{wackerrew,
Willenberg} already at $100$ K. The stable THz gain in tilted
magnetic field, by contrast, declines slowly and has similar
magnitudes as the usual Bloch gain also at room temperature.
Secondly, the large-signal response near the separatrix can be quite
different from the case of linear response. In the absence of
scattering the electron dynamics can even become chaotic.
Nevertheless, we found that the cyclotron gain in conditions of PDC
can still exist also for reasonably large amplitudes of the probe
field $E_\omega>2E_{cr}$.

We thank Feo Kusmartsev and Erkki Thuneberg for a constant
encouragement of this activity. This research was partially
supported by V\"{a}is\"{a}l\"{a} Foundation, University of Oulu, and
AQDJJ Programme of European Science Foundation.

\end{document}